\documentclass[12pt,a4paper]{article}
\usepackage{mathrsfs}
\usepackage{amsmath}
\usepackage{dsfont}
\usepackage{amsfonts}
\usepackage{amssymb}
\usepackage{graphicx}
\usepackage{txfonts}
\DeclareMathOperator*{\argmax}{\arg\!\max}
\newtheorem{definition}{Definition}

\title{Quantum signaling game}
\author{Piotr Fr\c{a}ckiewicz\\
\small Institute of Mathematics, Pomeranian University\\ \small 76-200 S\l upsk, Poland\\ \small P.Frackiewicz@impan.gov.pl}
\begin{document}
\maketitle
\begin{abstract}
We present a quantum approach to a signaling game; a special kind
of extensive
 games of incomplete information. Our model is based on quantum schemes for games in strategic
 form where players perform unitary operators on their own
 qubits of some fixed initial state and the payoff function is
 given by a measurement on the resulting final state. We show that the quantum game induced by our scheme coincides
 with a signaling game as a special case and outputs nonclassical
 results in general. As an example, we consider a quantum
 extension of the signaling game in which the chance move is a three-parameter unitary operator whereas the players' actions
 are equivalent to
 classical ones. In this case, we study the game in terms of Nash
 equilibria and refine the pure Nash equilibria adapting to the quantum
 game the
 notion of a weak perfect Bayesian equilibrium.

\end{abstract}
\section{Introduction}
The fifteen-year period of the development of quantum games has brought
some of the ideas that tell us how special extensive form games
might be played in the quantum domain, for example, the quantum model
of Stackelberg duopoly \cite{lo} or games with multiple rounds
\cite{gutowski}. However, the previous results do not explain
(even in a simple two-stage quantum game) how to identify
behavioral strategies, information sets and other terms connected
with extensive game theory.  In our recent papers
\cite{fracorimperfect, fracor1} we have proposed a way of quantizing
extensive games without chance moves through their normal
representation which covers not only two-stage extensive games but
also more complex games, including games with imperfect
information. In this paper, with the use of a signaling game, we
generalize our idea by allowing a chance mover to perform a
quantum operation.

The key feature of our research is the study of the extensive structure of
the quantum scheme so that we are able to introduce the notion of
perfect Bayesian equilibrium--a~Nash equilibrium refinement for
games of incomplete information. For convenience, we consider the
case where the chance mover and the players are equipped with
unitary operations. Thus, we assume that the only interaction with
the environment is by a quantum measurement. Certainly, a more
general scheme could be constructed. According to \cite{benjamin,
zhang}, the most natural generalization is to allow the players to
use general quantum operations, i.e., trace-preserving,
completely-positive maps. Our aim is not to construct the most
general scheme but to show that quantum game theory can be
developed by applying advanced terms from classical game theory.

 To make the paper self-contained, we begin
with recalling the notion of a signaling game and perfect Bayesian
equilibrium.
\section{Signaling game}
\begin{figure}[t]
\centering
\includegraphics[scale=0.6]{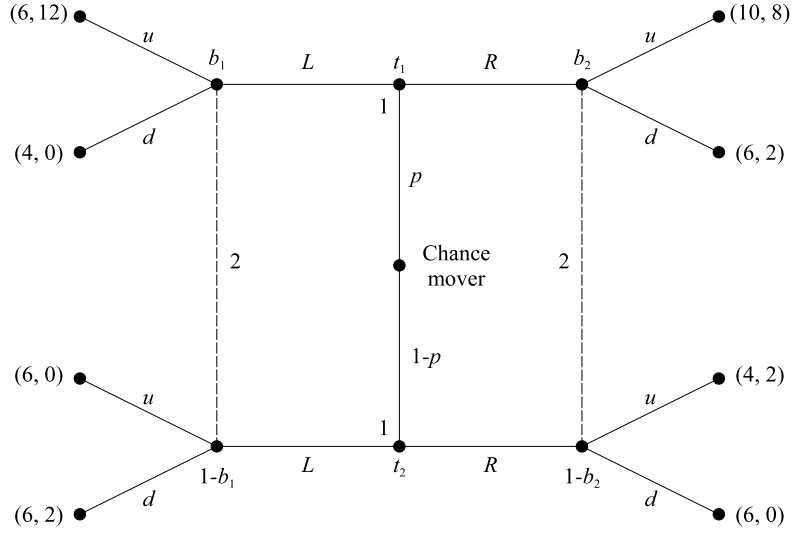}
\caption{A signaling game.\label{figure1}}
\end{figure}

The signaling game that we are going to study was introduced by
In-Koo Cho and David M Kreps in \cite{kreps}. The game begins
with a chance move that determines the type of player 1. After
player 1 is informed about her type, she chooses her action. Then
player 2 observes this action and moves next. The extensive form
of such a game is illustrated in figure~\ref{figure1}.

In this game each player has got two information sets--points of
the game that describe the player's knowledge about previous actions
chosen in the game. Player 1's information sets are
represented by single nodes $t_{1}$ and $t_{2}$ since player 1
knows exactly her type. On the other hand, player 2's
information sets are determined by the actions of player~1. They
are represented by the nodes connected by dashed lines that follow
player 1's actions. These information sets point out that player 2
learns about an action chosen by player 1. She does not know,
however, the type of player 1. This lack of knowledge is a key
feature of a signaling game.  The only way for player 2 to find
out player 1' type is to analyse her chosen actions that might
be a signal about this type.
\paragraph{Solution concepts for a signaling game} One of the
 most commonly used solution concepts
for noncooperative games is a Nash equilibrium \cite{nash} (see
also \cite{peters}). It is a strategy profile such that no player
gains by unilateral deviation from the equilibrium strategy. The
formal definition of a pure Nash equilibrium for a game in
strategic form is as follows.

Let $\left(N, \{S_{i}\}_{i\in N}, \{u_{i}\}_{i\in N}\right)$ be a
game in strategic form, where $N = \{1,2,\dots, n\}$, $n\in
\mathbb{N}$ is the set of players, $S_{i}$ is the set of
strategies of player $i\in N$ and $u_{i}\colon S_{1}\times
S_{2}\times \dots \times S_{n} \to \mathbb{R}$ is the payoff
function of player $i$ that assigns for every strategy profile
$(s_1,s_{2},\dots, s_n)$ payoff $u_i(s_1,s_{2},\dots, s_{n})$.
\begin{definition}\label{nashdef}
A profile of strategies $(s^{*}_1, s^{*}_2,\dots, s^{*}_{n})$ is a
pure Nash equilibrium in a~strategic game\\ $\left(N,
\{S_{i}\}_{i\in N}, \{u_{i}\}_{i\in N}\right)$ if for each player
$i \in N$ and for all $s_i \in S_i$
\begin{equation}\label{nash}
u_{i}(s^{*}_i, s^{*}_{-i}) \geq u_i(s_i, s^{*}_{-i}),~~
\mbox{where}~~ s^*_{-i} = (s^{*}_1,\dots, s^{*}_{i-1},
s^{*}_{i+1},\dots, s^{*}_{n}).
\end{equation}
\end{definition}

A Nash equilibrium is treated as a necessary condition for a
strategy profile to be a reasonable solution of any noncooperative
game and it may be very useful for determining a probable outcome
of a strategic game if the number of Nash equilibria is quite low.
However, the significance of a Nash equilibrium may decline if one
considers a game in extensive form. An extensive game may have a
lot of Nash equilibria and/or some of the Nash equilibria may
include actions which are not optimal off the equilibrium path
(see for example \cite{peters,vega}). As an example, let us
consider an extensive game in figure~\ref{figure1} with $p =
\frac{1}{2}$. One way to find pure Nash equilibria in this game is
first to determine its normal form. It is a strategic game defined
by the number of players, all the possible strategies of the
extensive game and the payoffs corresponding to the strategy
profiles. We recall that a player's strategy in an extensive game
is a function assigning an action to each information set of the
player. As a result, the normal form with $p=\frac{1}{2}$ is as
follows:
\begin{equation}\label{normalform}
\bordermatrix{&uu & ud & du & dd \cr
                LL& (6,6) & (6,6) & (5,1) & (5,1) \cr
                LR& (5,7) & (6,6) & (4,1) & (5,0) \cr
                RL& (8,4) & (6,1) & (8,5) & (6,2) \cr
                RR& (7,5) & (6,1) & (7,5) & (6,1)},
\end{equation}
where, for example, strategy $LR$ means that player 1 plays action
$L$ at her first information set and $R$ at the second one, and
$uu$ means that player 2 chooses action $u$ at both her
information sets. Using the system of inequalities (\ref{nash}),
pure Nash equilibria in (\ref{normalform}) are then $(RL,du)$ and
$(LL,ud)$. Although it might seem that these two equilibria are
equally likely scenarios of the game, it is supposed that
equilibrium strategy $ud$ will not be chosen by a rational player.
It follows from non-optimal action $d$ off the equilibrium path.
Indeed, note that action $d$ is strictly dominated by action $u$
at the right information set of player 2, i.e., it always gives a
worse outcome than action $u$ at this information set. Thus, in
case player 2 specifies her action at the right information set,
she ought to choose $u$ instead.

Equilibrium refinements can exclude Nash equilibria containing
non-optimal actions. For the signaling game it is sufficient to
consider a (weak) perfect Bayesian equilibrium, where, for our
needs, we restrict ourselves to pure strategies. Following Kreps
and Wilson~\cite{kreps2}, let us first define an assessment to be
a pair $(s,\mu)$ of a pure strategy profile $s$ and a belief
system $\mu$, i.e., a map that assigns to each information set a
probability distribution over the nodes of this information set.
Thus, in our example in figure~(\ref{figure1}), player 2's beliefs
are probability distributions $(b_{1}, 1- b_{1})$ and $(b_{2}, 1-
b_{2})$. In turn, player 1's beliefs assign to her decision nodes
a probability equal to 1. Now, for any node $x$ from an information
set $h$ let $P(x)$ denote the probability that $x$ is reached
given $s$ and chance moves, if any.
\begin{definition}
An assessment $(s,\mu)$ is Bayesian consistent if belief $\mu(x)$
at node $x$ is equal to $P(x)/\sum_{x\in h}P(x)$ for all $h$ for
which $\sum_{x\in h}P(x)>0$ and all $x \in h$.
\end{definition}
Denote by $u_{i}(a|s,x)$ the expected payoff of player $i$ from
playing action $a$, conditional on being at node $x$ and strategy
profile $s$. Then, $\sum_{x\in h}\mu(x)u_{i}(a|s,x)$ is the
expected payoff of player $i$ from playing the action $a$,
conditional on being at information set $h$.
\begin{definition}
An assessment $(s,\mu)$ in an extensive game is sequentially
rational if for each player $i \in N$, each information set $h$ of
this player and action $a$ from the set $A(h)$ of available
actions at $h$, if $a$ is consistent with $s_{i}$ then
\begin{equation}\label{conditionsequential}
\sum_{x\in h}\mu(x)u_{i}(a|s,x) = \max_{a'\in A(h)}{\sum_{x\in
h}\mu(x)u_{i}(a'|s,x)}.
\end{equation}
\end{definition}

In other words, condition (\ref{conditionsequential}) requires for
each player $i$ that action $a$ prescribed by strategy $s_{i}$ is
optimal given $s=(s_{i})_{i\in \mathbb{N}}$ and beliefs $\mu$. The
two definitions above allow one to formulate the following
equilibrium refinement:
\begin{definition}
An assessment $(s,\mu)$ in an extensive game is a (weak) perfect
Bayesian equilibrium if it is sequentially rational and Bayesian
consistent.
\end{definition}

Let us consider, for example, Nash equilibrium $(LL,ud)$ with a
view to a perfect Bayesian equilibrium. If probability
distribution $(p,1-p) = (1/2, 1/2)$ determines the chance moves,
the Bayesian consistency on player 2's beliefs requires $b_{1} =
1/2$ whereas beliefs $(b_{2}, 1-b_{2})$ are not forced by the
consistency requirement. In this case action $u$ at the left
information set is optimal. However, given arbitrary beliefs
$(b_{2}, 1-b_{2})$, action $d$ at the right information set gives
player 2 a lower payoff than action $u$. As a consequence, the
profile $(LL,ud)$ is not a perfect Bayesian Nash equilibrium.
A similar analysis would show that the profile $(RL,du)$ together
with player 2' beliefs $b_{1} = 0$ and $b_{2}=1$ satisfies
Bayesian consistency and sequential rationality.

\section{Quantum model for a signaling game}
In paper \cite{fracor1} we introduced a model for describing
extensive games, where we focused on the normal form and studied
Nash equilibria in the resulting quantum game. Here, we are going to
justify our scheme with respect to the dynamic nature of an
extensive game.

\paragraph{Motivation for the model construction} Let us consider the generalized Eisert-Wilkens-Lewenstein (EWL)
quantum approach to an $n$-player strategic game with two-element
strategy sets \cite{benjamin} (we encourage readers who are not
familiar with the EWL scheme to first see \cite{eisert}).
According to an alternative notation for the EWL scheme introduced
in \cite{eisert2} and generalized in \cite{benjamin}, the quantum
protocol is defined by 4-tuple
\begin{equation}\label{ewl}
\left\{\mathcal{H}, |\Psi_{\mathrm{in}}\rangle, \mathsf{SU}(2),
\{M_{i}\}_{i\in \{1,2,\dots,n\}}\right\},
\end{equation}
where the components specify the game in the following way:
\begin{itemize}
\item $\mathcal{H}$ is a Hilbert space $(\mathbb{C}^2)^{\otimes
n}$ with basis $\left\{|\Psi_{x_{1}x_{2}\dots
x_{n}}\rangle\right\}_{x_{1},x_{2},\dots,x_{n} \in \{0,1\}}$
defined for all $(x_{1},x_{2},\dots,x_{n}) \in \{0,1\}^n$ by the
formula
\begin{equation}
|\Psi_{x_{1}x_{2}\dots x_{n}}\rangle = \frac{|x_{1}x_{2}\dots
x_{n}\rangle + \mathrm{i}|\overline{x}_{1}\overline{x}_{2}\dots
\overline{x}_{n}\rangle}{\sqrt{2}},
\end{equation}
where $\overline{x}_{i}$ is the negation of $x_{i}$.  \item
$|\Psi_{\mathrm{in}}\rangle$ is called the initial state, and
$|\Psi_{\mathrm{in}}\rangle = |\Psi_{00\dots 0}\rangle$. \item
$\mathsf{SU}(2)$ defines the unitary operators available for each
player. The matrix representation of the operators from
$\mathsf{SU}(2)$ (with respect to the computational basis) can be
written as follows:
\begin{equation}\label{unitary1}
U_{i}(\theta_{i},\alpha_{i},\beta_{i}) = \left(\begin{array}{cc}
\mathrm{e}^{\mathrm{i}\alpha_{i}}\cos{\frac{\theta_{i}}{2}} &
\mathrm{i}\mathrm{e}^{\mathrm{i}\beta_{i}}\sin{\frac{\theta_{i}}{2}}\\
\mathrm{i}\mathrm{e}^{-\mathrm{i}\beta_{i}}\sin{\frac{\theta_{i}}{2}} &
\mathrm{e}^{-\mathrm{i}\alpha_{i}}\cos{\frac{\theta_{i}}{2}}
\end{array}\right).
\end{equation}
\item $M_{i}$ for each $i\in \{1,\dots,n\}$ is an observable given
by the formula
\begin{equation}
M_{i} = \sum_{x_{1},x_{2},\dots,x_{n} \in
\{0,1\}}m^{i}_{x_{1}x_{2}\dots x_{n}}|\Psi_{x_{1}x_{2}\dots
x_{n}}\rangle \langle \Psi_{x_{1}x_{2}\dots x_{n}}|.
\end{equation}
The measurement is performed on the final state
$|\Psi_{\mathrm{f}}\rangle = \bigotimes^{n}_{i=1}U_{i}(\theta_{i},
\alpha_{i}, \beta_{i})|\Psi_{\mathrm{in}}\rangle $. The possible
outcomes $m^i_{x_{1}x_{2}\dots x_{n}} \in \mathbb{R}$ of the
measurement correspond to player $i$'s payoffs.
\end{itemize}
It turns out that we can adapt scheme~(\ref{ewl}) for any
extensive game with two actions at each information set. Since
operators $U_{i}(\theta_{i},0,0)$ represent classical moves in the
EWL scheme for a $2\times 2$ bimatrix game, it is natural to assume
that they correspond to classical moves in any quantum game
defined by the generalized scheme. The argumentation is as
follows. The final state $|\Psi_{\mathrm{f}}\rangle$ after each
player performs her unitary operator $U_{i}(\theta_{i}, 0, 0)$ is
as follows:
\begin{align}
|\Psi_{\mathrm{f}}\rangle &=\bigotimes^{n}_{i=1}U_{i}(\theta_{i},
0,
0)|\Psi_{\mathrm{in}}\rangle \nonumber\\
& = \bigotimes^{n}_{i=1}\left(\cos{\frac{\theta_{i}}{2}}|0\rangle
+ \mathrm{i}\sin{\frac{\theta_{i}}{2}}|1\rangle\right) +
\mathrm{i}\bigotimes^n_{i=1}\left(\mathrm{i}\sin{\frac{\theta_{i}}{2}}|0\rangle+\cos{\frac{\theta_{i}}{2}}|1\rangle
\right)\nonumber\\
&=\sum_{x_{1},x_{2},\dots,x_{n}\in
\{0,1\}}\mathrm{i}^{\sum^n_{j=1}x_{j}}\cos{\left(\frac{x_{1}\pi -
\theta_{1}}{2}\right)}\cdots\cos{\left(\frac{x_{n}\pi -
\theta_{n}}{2}\right)}|\Psi_{x_{1},\dots,x_{n}}\rangle.
\end{align}
Let us denote by $x_{i} \in \{0_{i}, 1_{i}\}$ an action at the
$i$th information set and by
\begin{equation}
\begin{split}&P_{x_{i}}\coloneqq \sum_{x_{k}\in\{0,1\}, k\ne
i}|\Psi_{x_{1}x_{2}\dots x_{n}}\rangle \langle
\Psi_{x_{1}x_{2}\dots x_{n}}|\\
&P_{x_{i}x_{j}}\coloneqq \sum_{x_{k}\in\{0,1\}, k\ne
i,j}|\Psi_{x_{1}x_{2}\dots x_{n}}\rangle \langle
\Psi_{x_{1}x_{2}\dots x_{n}}|\\
&~~~~\vdots \\
&P_{x_{1}x_{2}\dots x_{n}} \coloneqq |\Psi_{x_{1}x_{2}\dots
x_{n}}\rangle \langle \Psi_{x_{1}x_{2}\dots x_{n}}|\end{split}
\end{equation}
the projectors onto the respective subspaces of
$(\mathbb{C}^{2})^{\otimes n}$. Let us assign to each action
$x_{i}$ projection $P_{x_{i}}$ of the state vector
$|\Psi_{\mathrm{f}}\rangle$. Then, $\langle
\Psi_{\mathrm{f}}|P_{0_{i}}|\Psi_{\mathrm{f}}\rangle =
\cos^2{\frac{\theta_{i}}{2}}$ and $\langle
\Psi_{\mathrm{f}}|P_{1_{i}}|\Psi_{\mathrm{f}}\rangle =
\sin^2{\frac{\theta_{i}}{2}}$. Taking $p \coloneqq
 \cos^2{\frac{\theta_{i}}{2}}$, we obtain a probability distribution
 over the actions equivalent to one given by a classical behavioral strategy $b= (p, 1-p)$. Thus, in particular, $U_{i}(0,0,0)$ and $U_{i}(\pi,0,0)$ represent pure actions. In general, let us assign to a sequence of
 actions $x_{i_{1}}, x_{i_{1}}, \dots, x_{i_{k}}$ the product of
 projections $\prod^k_{j=1}P_{x_{i_{j}}} = P_{x_{i_{1}}x_{i_{1}}\dots
 x_{i_{k}}}$. Then $\langle
\Psi_{\mathrm{f}}|P_{x_{i_{1}}x_{i_{1}}\dots
 x_{i_{k}}}|\Psi_{\mathrm{f}}\rangle = \prod^k_{j=1}\cos^2{\frac{x_{i_j}\pi -
\theta_{i_j}}{2}}$ and this corresponds to the product of
probabilities given by applying a sequence $(b_{i_{j}})$ of
classical behavioral strategies $b_{i_{j}} =
\left(\cos^2{\frac{\theta_{i_{j}}}{2}},\sin^2{\frac{\theta_{i_{j}}}{2}}\right)$
at the $i_{j}$th information set, where $j = 1,\dots,k$.

\begin{figure}[t]
\centering
\includegraphics[scale=0.6]{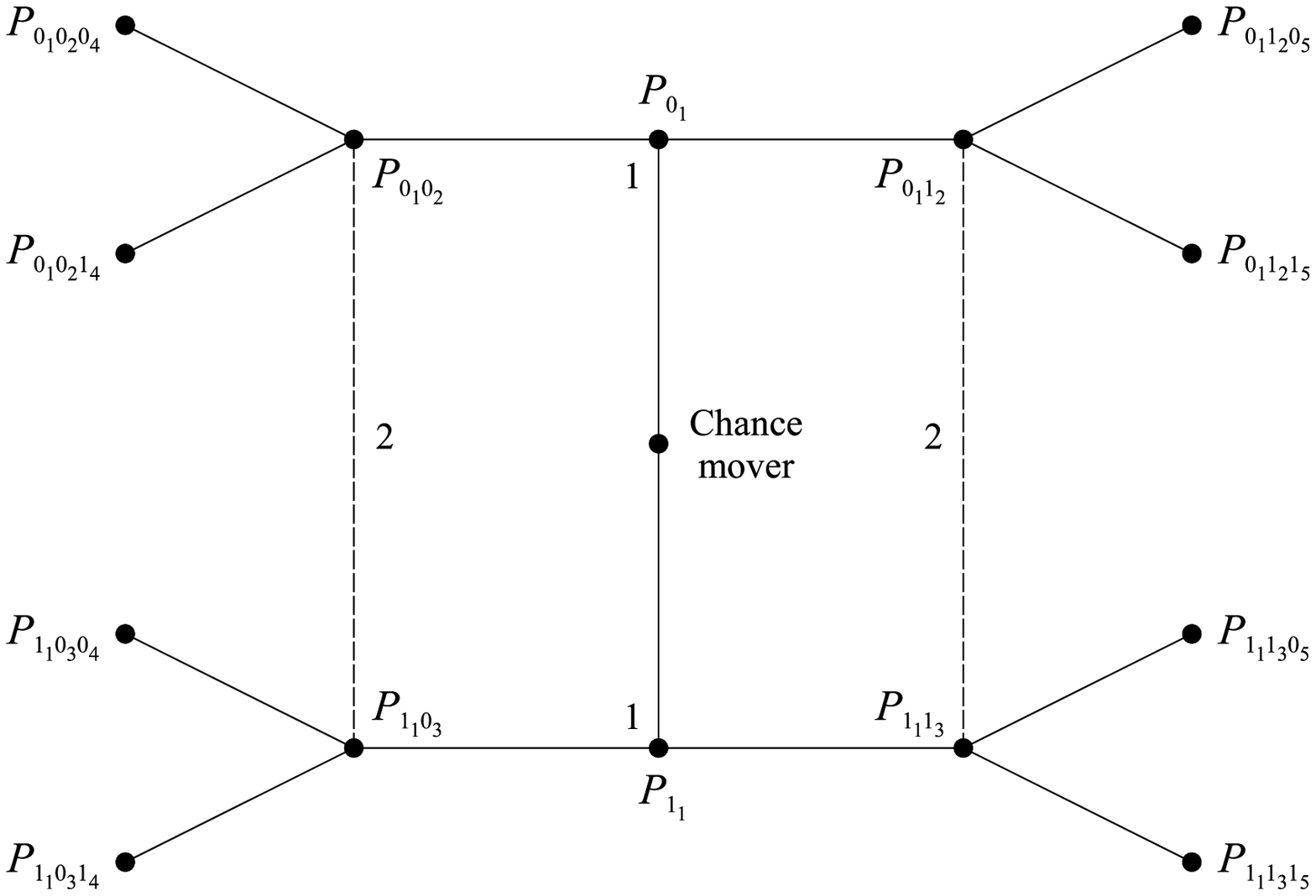}
\caption{Projective measurements corresponding to particular game
nodes. \label{figure2}}
\end{figure}

As a result, we have obtained the procedure how to describe an
extensive game in terms of the mathematical methods of quantum
information. At the same time, we have obtained a scheme that places an
extensive game in quantum domain whenever the set of unitary
operators of at least one player is $\mathsf{SU}(2)$.
\paragraph{Quantum
model} A detailed description of the quantum scheme for a game in
figure~\ref{figure1} is as follows. This is a 6-tuple
\begin{equation}\label{6tuple}
\left(\mathcal{H}, N\cup\{C\}, |\Psi_{\mathrm{in}}\rangle,\xi,
\mathsf{SU}(2), \{M_{i}\}_{i\in N} \right),
\end{equation}
where
\begin{itemize}
\item components $\mathcal{H}$ and $|\Psi_{\mathrm{in}}\rangle$
are the special case of those from (\ref{ewl}) for a Hilbert space
$(\mathbb{C}^2)^{\otimes 5}$; \item $N = \{1,2\}$ is a set of
players and $C$ is a chance mover; \item $\mathsf{SU}(2)$
specifies the players' and the~chance mover's actions. It is
assumed that a unitary operation performed by the chance mover is
known to the players; \item $\xi$ is a map that relates qubits to
players and the chance mover. It is a map $\xi\colon
\{1,2,\dots,5\} \to N\cup \{C\}$ given by formula
\begin{equation}
\xi(j) = \begin{cases} C & \mbox{if}~ j=1 \\ 1 & \mbox{if}~ j \in
\{2,3\} \\ 2 & \mbox{if}~ j \in \{4,5\},
\end{cases}
\end{equation}
 that assigns to each index $j\in \{1,\dots,5\}$ of $x_{j}$ in
 $|\Psi_{x_{1}x_{2}x_{3}x_{4}x_{5}}\rangle$ a player or the chance
mover; \item $M_{i}$ is an observable that describes a measurement
on the final state $|\Psi_{\mathrm{f}}\rangle$,
\begin{multline}\label{observable}
M_{i} = m^i_{1}P_{0_{1}0_{2}0_{4}} + m^i_{2}P_{0_{1}0_{2}1_{4}} +
m^i_{3}P_{1_{1}0_{3}0_{4}}  + m^i_{4}P_{1_{1}0_{3}1_{4}}\\  +
m^i_{5}P_{0_{1}1_{2}0_{5}} + m^i_{6}P_{0_{1}1_{2}1_{5}} +
m^i_{7}P_{1_{1}1_{3}0_{5}} + m^i_{8}P_{1_{1}1_{3}1_{5}}
\end{multline}
Then the average value $E_{i}$ of measurement $M_{i}$,
\begin{equation} E_{i} = \langle
\Psi_{\mathrm{f}}|M_{i}|\Psi_{\mathrm{f}}\rangle~~ \mbox{for}~~
|\Psi_{\mathrm{f}}\rangle =
\bigotimes^{5}_{i=1}U_{i}(\theta_{i},\alpha_{i},\beta_{i})|\Psi_{00000}\rangle
\end{equation} determines a payoff for player $i \in N$.
\end{itemize}
Thus, the quantum model for the signaling game in
figure~\ref{figure1} requires a five-qubit state. The chance
mover's action is represented by a unitary operation on the first
qubit. In turn, a unitary operation $U_{2}\otimes U_{3}$ on the
second and third qubit, and a unitary operation $U_{4}\otimes
U_{5}$ on the fourth and fifth one are player 1's and player 2's
strategies, respectively. The form of observable~(\ref{observable})
is based on our motivation for the scheme construction. Following
this line of thought, and then the link between projections and
actions as it is given in figure~\ref{figure2}, each term in
$M_{i}$ corresponds to measurement that the state of the game is
in the respective end node.
\section{A signaling game with a quantum chance mover}
In the literature of quantum games one can find many examples that
show the advantages of quantum strategies over classical ones. The
same could be done for the quantum signaling game if, for example,
one player's strategy set were extended to the full range of
unitary operators. We are going to consider another case where the
players' actions are still classical ones, i.e., they are in the
form of $U(\theta,0,0)$, and the full set $\mathsf{SU}(2)$ is
available only for the chance mover. In this case, we obtain an
interesting example, where the normal form of the resulting
quantum game has the same dimension as the classical game. As it
is shown below, this feature makes the classical and quantum game
easy to compare. Moreover, the fact that the players are equipped
with operators $U(\theta,0,0)$ enables us, easily, to
refine Nash equilibria in the quantum game by using the notion of
perfect Bayesian equilibrium.

More precisely, let us consider 6-tuple~(\ref{6tuple}) in which
the set $\mathsf{SU}(2)$ is available only for the chance mover,
and the players are equipped with the set $\{U(\theta)\coloneqq
U(\theta,0,0)\colon \theta \in [0,\pi]\}$. The possible
measurement $M_{i}$'s outcomes $m^i_{j}$, for $j=1,\dots,8$ of
measurement $M_{i}$ correspond to payoffs from the game in
figure~(\ref{figure1}), i.e.,
\begin{equation}\label{eqn2}
\begin{split}(m^{1}_{1},m^{2}_{2}) = (6,12) \quad
(m^{1}_{2},m^{2}_{2}) = (4,0) \quad (m^{1}_{3},m^{2}_{3}) = (6,0)
\quad (m^{1}_{4},m^{2}_{4}) = (6,2);\\ (m^{1}_{5},m^{2}_{5}) =
(10,8) \quad (m^{1}_{6},m^{2}_{6}) = (6,2) \quad
(m^{1}_{7},m^{2}_{7}) = (4,2) \quad (m^{1}_{8},m^{2}_{8}) =
(6,0).\end{split}
\end{equation}
Let us assume that the chance mover specifies $U_{1}(\pi/2, \pi/6,
\pi/3)$ as her move. We recall that a chance mover's action
$U_{1}(\pi/2,0,0)$ corresponds to probability distribution
$(1/2,1/2)$ over her classical actions and non zero coordinates
$\alpha_{1}$ and $\beta_{1}$ in $U_{1}(\theta_{1}, \alpha_{1},
\beta_{1})$ place the game into the quantum domain. The final
state after the first and second player specify
$U_{2}(\theta_{2})\otimes U_{3}(\theta_{3})$ and $U(\theta_{4})
\otimes U(\theta_{5})$, respectively, is in the form
\begin{equation}
|\Psi_{\mathrm{f}}\rangle = \left(U_{1}\left(\frac{\pi}{2},
\frac{\pi}{6}, \frac{\pi}{3}\right) \otimes
U_{2}(\theta_{2})\otimes U_{3}(\theta_{3}) \otimes
U_{4}(\theta_{4})\otimes
U_{5}(\theta_{5})\right)|\Psi_{00000}\rangle.
\end{equation}
Let us calculate the expected values $E_{1}$ and $E_{2}$ for each
$(\theta_{2}, \theta_{3}, \theta_{4}, \theta_{5}) \in
\{0,\pi\}^4$; values corresponding to pure strategy profiles. For
example, quadruple $(0, 0, \pi, \pi)$ implies the following final
state:
\begin{equation}
|\Psi_{\mathrm{f}}\rangle =
-\frac{\sqrt{6}}{4}|\Psi_{00011}\rangle
-\frac{\sqrt{2}}{4}|\Psi_{11100}\rangle -
\frac{\sqrt{2}\mathrm{i}}{4}|\Psi_{10011}\rangle +
\frac{\sqrt{6}\mathrm{i}}{4}|\Psi_{01100}\rangle.
\end{equation}
Then, the pair $(\langle
\Psi_{\mathrm{f}}|M_{1}|\Psi_{\mathrm{f}}\rangle,
\langle\Psi_{\mathrm{f}}|M_{2}|\Psi_{\mathrm{f}}\rangle)$ equals
$(6.5,3.5)$. By calculating the average values of measurements $M_{1}$
and $M_{2}$ for the other quadruples we obtain the following
bimatrix:
\begin{equation}\label{normalquantumform}
\bordermatrix{&U_{4}(0)\otimes U_{5}(0) & U_{4}(0)\otimes
U_{5}(\pi) & U_{4}(\pi)\otimes U_{5}(0) & U_{4}(\pi)\otimes
U_{5}(\pi) \cr
                U_{2}(0)\otimes U_{3}(0)& (6,5.25) & (7.25,7.75) & (5.25,1) & (6.5,3.5) \cr
                U_{2}(0)\otimes U_{3}(\pi)& (5.75,5.75) & (7.5,7.75) & (5,1) & (6.75,3) \cr
                U_{2}(\pi)\otimes U_{3}(0)& (6.75,3) & (5,1) & (7.5,7.75) & (5.75,5.75) \cr
                U_{2}(\pi)\otimes U_{3}(0)& (6.5,3.5) & (5.25,1) & (7.25,7.75) &
                (6,5.25)}.
\end{equation}
As a result, quantum scheme (\ref{6tuple}) provides the players
with a quite different bimatrix compared with (\ref{normalform}).
In particular, the classical game and the quantum counterpart have
two pure Nash equilibria but differ in payoff outcomes. Indeed, in
contrast to the classical case, profiles $\bigl((U_{2}(0)\otimes
U_{3}(\pi)),(U_{4}(0)\otimes U_{5}(\pi))\bigr)$ and
$\bigl((U_{2}(\pi)\otimes U_{3}(0)),(U_{4}(\pi)\otimes
U_{5}(0))\bigr)$ are Nash equilibria with the same payoff outcome
$(7.5,7.75)$.


\paragraph{Perfect Bayesian-type equilibria} Let us study profile $\bigl((U_{2}(\pi)\otimes U_{3}(0)),(U_{4}(\pi)\otimes U_{5}(0))\bigr)$.
 According to definition~\ref{nashdef}, no player gains by unilaterally deviating from unitary operator $U(\pi)\otimes U(0)$. It turns out that it can be said more about the profile in terms of Bayesian consistency and sequential rationality. First, let us carry out perfect Bayesian equilibrium analysis for player 1. Following figure~\ref{figure2}, the probability that the game reaches the upper node given chance move $U_{1}(\pi/2, \pi/6, \pi/3)$ and player 2's strategy $(U_{4}(\pi)\otimes U_{5}(0))$ equals $\langle\Psi_{1}|P_{0}|\Psi_{1}\rangle =
 3/4$ where
 \begin{align}
 |\Psi_{1}\rangle &= \left(U_{1}\left(\frac{\pi}{2}, \frac{\pi}{6}, \frac{\pi}{3}\right) \otimes \mathds{1} \otimes \mathds{1} \otimes U_{4}(\pi)\otimes U_{5}(0)\right)|\Psi_{00000}\rangle \nonumber\\
 &=\frac{\sqrt{6}\mathrm{i}}{4}|\Psi_{00010}\rangle + \frac{\sqrt{2}\mathrm{i}}{4}|\Psi_{11101}\rangle - \frac{\sqrt{2}}{4}|\Psi_{10010}\rangle +
 \frac{\sqrt{6}}{4}|\Psi_{01101}\rangle,
 \end{align}
and $\mathds{1}$ means the identity operator on $\mathbb{C}^2$.
Since player 1's information sets are singletons, reaching the
upper node is equivalent to reaching the corresponding information
set, so her belief of being at the upper node attaches probability
1 to this node. Therefore, after player 1 learns that the state of
the chance mover's qubit corresponds to $P_{0_{1}}$, she believes
that with probability 1 faces the following state:
 \begin{equation}\label{statep0}
 |\Psi'_{1}\rangle  = \frac{P_{0_{1}}|\Psi_{1}\rangle}{\sqrt{\langle \Psi_{1}|P_{0_{1}}|\Psi_{1}\rangle}} = \frac{\sqrt{2}\mathrm{i}}{2}|\Psi_{00010}\rangle +
 \frac{\sqrt{2}}{2}|\Psi_{01101}\rangle.
 \end{equation}
Denote by $|\Psi''_{1}\rangle$ the state obtained when player 1
performs unitary strategy $U_{2}(\theta_{2})\otimes
U_{3}(\theta_{3})$ on $|\Psi'_{1}\rangle$. Then
\begin{align}
|\Psi''\rangle &= \bigl(\mathds{1}\otimes U_{2}(\theta_{2}) \otimes U_{3}(\theta_{3}) \otimes \mathds{1} \otimes \mathds{1} \bigr)|\Psi'_{1}\rangle\nonumber\\
&=\frac{\sqrt{2}}{2}\sum_{x_{2},x_{3} \in
\{0.1\}}\mathrm{i}^{x_{2}+x_{3}}c(x_{2},x_{3})(\mathrm{i}|\Psi_{0x_{2}x_{3}10}\rangle
+ |\Psi_{0\overline{x}_{2}\overline{x}_{3}01}\rangle),
\end{align}
where $c(x_{2},x_{3}) =
\cos{\left(\frac{x_{2}\pi-\theta_{2}}{2}\right)}\cos{\left(\frac{x_{3}\pi-\theta_{3}}{2}\right)}$.
Then, $U_{2}(\pi) \otimes U_{3}(0)$ is optimal given the belief
(\ref{statep0}) about the quantum state if
\begin{equation}\label{argmax1}
U_{2}(\pi) \otimes U_{3}(0) \in \argmax_{U_{2}(\theta_{2}) \otimes
U_{3}(\theta_{3})}\langle \Psi''_{1}|M_{1}|\Psi''_{1}\rangle.
\end{equation}
Indeed,
\begin{equation}
\max_{U_{2}(\theta_{2}) \otimes U_{3}(\theta_{3})}\langle
\Psi''_{1}|M_{1}|\Psi''_{1}\rangle =
\max_{U_{2}(\theta_{2})}\left(8-3\cos^{2}{\frac{\theta_{2}}{2}}\right)
= 8.
\end{equation}
Thus, condition (\ref{argmax1}) is satisfied.

Similar computation for the case when player 1 learns that the
state of the chance mover's qubit corresponds to $P_{1_{1}}$
proves that $U_{2}(\pi) \otimes U_{3}(0)$ is also optimal on state
\begin{equation}
\frac{P_{1_{1}}|\Psi_{1}\rangle}{\sqrt{\langle
\Psi_{1}|P_{1_{1}}|\Psi_{1}\rangle}} =
\frac{\sqrt{2}\mathrm{i}}{2}|\Psi_{11101}\rangle -
\frac{\sqrt{2}}{2}|\Psi_{10010}\rangle.
\end{equation}
As a result, player 1's strategy $U_{2}(\pi) \otimes U_{3}(0)$ is sequentially-type rational given her beliefs.

Let us consider now player 2's strategy $U_{4}(\pi)\otimes U_{5}(0)$ in the terms of perfect Bayesian equilibrium. The state after the chance mover and player 1 use operators $U_{1}\left(\frac{\pi}{2}, \frac{\pi}{6}, \frac{\pi}{3}\right)$ and $U_{2}(\pi)\otimes U_{3}(0)$, respectively, is as follows
\begin{align}
 |\Psi_{2}\rangle &= \left(U_{1}\left(\frac{\pi}{2}, \frac{\pi}{6}, \frac{\pi}{3}\right) \otimes U_{2}(\pi) \otimes U_{3}(0) \otimes \mathds{1} \otimes \mathds{1}\right)|\Psi_{00000}\rangle \nonumber\\
 &=\frac{\sqrt{6}\mathrm{i}}{4}|\Psi_{01000}\rangle + \frac{\sqrt{2}\mathrm{i}}{4}|\Psi_{10111}\rangle - \frac{\sqrt{2}}{4}|\Psi_{11000}\rangle + \frac{\sqrt{6}}{4}|\Psi_{00111}\rangle.
\end{align}
Then the probability that the left information set is reached is
equal to $\langle \Psi_{2}|P_{0_{1}0_{2}}|\Psi_{2}\rangle +
\langle \Psi_{2}|P_{1_{1}0_{3}}|\Psi_{2}\rangle = 1/2$. By
Bayesian consistency, player 2's beliefs of being at the upper and
lower node at the left information set are
\begin{equation}
\frac{\langle \Psi_{2}|P_{0_{1}0_{2}}|\Psi_{2}\rangle}{\langle \Psi_{2}|P_{0_{1}0_{2}}|\Psi_{2}\rangle + \langle \Psi_{2}|P_{1_{1}0_{3}}|\Psi_{2}\rangle}=\frac{3}{4} \quad \mbox{and} \quad \frac{\langle \Psi_{2}|P_{1_{1}0_{3}}|\Psi_{2}\rangle}{\langle \Psi_{2}|P_{0_{1}0_{2}}|\Psi_{2}\rangle + \langle \Psi_{2}|P_{1_{1}0_{3}}|\Psi_{2}\rangle}=\frac{1}{4}.
\end{equation}
As a consequence, specifying her beliefs, player 2 faces
post-measurement state
$P_{0_{1}0_{2}}|\Psi_{2}\rangle/\sqrt{\langle
\Psi_{2}|P_{0_{1}0_{2}}|\Psi_{2}\rangle} = |\Psi_{00111}\rangle$
with probability 3/4, and state
$P_{1_{1}0_{3}}|\Psi_{2}\rangle/\sqrt{\langle
\Psi_{2}|P_{1_{1}0_{3}}|\Psi_{2}\rangle} = -|\Psi_{11000}\rangle$
with probability 1/4. In other words, player 2 is faced with the
following mixed state
\begin{equation}\label{jakisstate}
\rho_{2} = \frac{3}{4}|\Psi_{00111}\rangle \langle \Psi_{00111}| + \frac{1}{4}|\Psi_{11000}\rangle \langle \Psi_{11000}|.
\end{equation}
Thus, mixed state~(\ref{jakisstate}) after player 2 uses her unitary operator $U_{4}(\theta_{4}) \otimes U_{5}(\theta_{5})$ takes the form
\begin{equation}
\rho'_{2} = \left(\mathds{1}^{\otimes 3} \otimes U_{4}(\theta_{4})
\otimes U_{5}(\theta_{5})\right)\rho_{2} \left(\mathds{1}^{\otimes
3} \otimes U_{4}(\theta_{4}) \otimes
U_{5}(\theta_{5})\right)^{\dag}
\end{equation}
and player 2's expected payoff is given by
$\mathrm{tr}(\rho'_{2}M_{2})$. In order to prove that, given her
beliefs, $U_{4}(\pi) \otimes U_{5}(0)$ is optimal for player 2 let
us determine $\argmax_{U_{4}(\theta_{4})\otimes
U_{5}(\theta_{5})}\mathrm{tr}(\rho'_{2}M_{2})$,
\begin{align}\label{ostatniresult}\nonumber
&\argmax_{U_{4}(\theta_{4})\otimes
U_{5}(\theta_{5})}\mathrm{tr}(\rho'_{2}M_{2}) \\&\qquad =
\argmax_{U_{4}(\theta_{4})\otimes
U_{5}(\theta_{5})}\mathrm{tr}\left(\sum_{x_{4},x_{5}\in
\{0,1\}}c^2(x_{4},x_{5})\left(\frac{3}{4}|\Psi_{001\overline{x}_{4}\overline{x}_{5}}\rangle
\langle \Psi_{001\overline{x}_{4}\overline{x}_{5}}|+
\frac{1}{4}|\Psi_{110x_{4}x_{5}}\rangle \langle
\Psi_{110x_{4}x_{5}}| \right) \right)\nonumber\\
&\qquad=\argmax_{U_{4}(\theta_{4})\otimes
U_{5}(\theta_{5})}\frac{19}{2}\sin{\frac{\theta_{4}}{2}} =
\{\pi\}\times [0,\pi].
\end{align}
Result (\ref{ostatniresult}) shows that $U_{4}(\pi) \otimes
U_{5}(0)$ is also sequentially-type rational given player 2's
beliefs at the left information set. In a similar way, we can prove
sequential-type rationality of $U_{4}(\pi) \otimes U_{5}(0)$ at
the right information set.

As a result, strategy profile $\bigl((U_{2}(\pi)\otimes U_{3}(0)),(U_{4}(\pi)\otimes U_{5}(0))\bigr)$ consists of strategies that are optimal with respect to unilateral deviation in both cases: when only the payoff measuerment is performed (a Nash equilibrium) and when a player performs the additional measurement before her move (sequential rationality). It can be shown that the other pure equilibrium given by bimatrix~(\ref{normalquantumform}) is also a perfect Bayesian-type equilibrium.

\section{Conclusion and further research}

The purpose of the research was to translate signaling games into
the formalism of quantum information and to examine how playing the
game would then change. We showed that there exists a quantum
approach to a signaling game that constitutes a generalization of
the classical game. In particular, we proved (with the use of
Eisert {\it et al} quantum scheme for strategic games) that the special
one-parameter unitary strategies are equivalent to classical moves
in the game and a broader range of unitary operators affects the
game. The key result of our work was to show that optimal strategy
analysis in quantum games can go beyond the concept of Nash
equilibrium. A player measuring the state after the other players
operate but before her move gives rise to a new solution concept
in quantum games that can be treated as a counterpart of a perfect
Bayesian equilibrium in classical game theory. It is worth noting
that there is more than one way to define a quantum counterpart
of a perfect Bayesian equilibrium consistent with the classical
term. For example, when solving optimization problems (\ref{argmax1})
and (\ref{ostatniresult}) it does not matter whether we maximize
over $U_{2}(\theta_{2}) \otimes U_{3}(\theta_{3})$ and
$U_{4}(\theta_{4}) \otimes U_{4}(\theta_{4})$ or over
$U_2(\theta_{2})$ and $U_{4}(\theta)$. However, it may have great
importance when the full set of unitary operators is also
available for players and may constitute independent subject of
research. Another interesting problem would be to specify how the
players' strategic positions change when one or both players are
provided with the set $\mathsf{SU}(2)$. In particular, studying
player 2' position in the game seems significant. In the classical
game, she is deprived of knowing the type of player 1 and we
suppose that the access to quantum strategies may improve player
2' strategic position. Finally, one may investigate the
relation between Nash equilibrium and perfect Bayesian
equilibrium. We suppose that, in contrast to the classical case, the
perfect Bayesian conditions in the way we have presented in the paper
may not imply Nash equilibrium. This would point out another
distinction between classical and quantum game theory.
\section*{Acknowledgments}
The author is very grateful to Prof. J. Pykacz from the Institute
of Mathematics, University of Gda\'nsk, Poland for very useful
discussions and great help in putting this paper into its final
form. The project was supported by the Polish National Science
Center under the project number DEC-2011/03/N/ST1/02940.

\end{document}